\documentclass[a4paper, 12pt]{article}
\usepackage{graphicx}  
\usepackage{amsmath, amsthm,  amsfonts, amscd,amssymb,latexsym}
\usepackage{multirow}
\bibstyle{plain}
\title{Estimating Bayesian networks for high-dimensional data with complex mean structure and random effects}

\author{Jessica Kasza, Gary Glonek, Patty Solomon}

\begin{document}

\maketitle{}
\begin{abstract}
The estimation of Bayesian networks given high-dimensional data, in particular gene expression data, has been the focus of much recent research.  Whilst there are several methods available for the estimation of such networks, these typically assume that the data  consist of independent and identically distributed samples. However, it is often the case that the available data  have a more complex mean structure plus additional components of variance, which must then be accounted for in the estimation of a Bayesian network. 
In this paper, score metrics that take account of such complexities are proposed for use in conjunction with score-based methods for the estimation of Bayesian networks. We propose firstly, a fully Bayesian score metric, and secondly, a metric inspired by the notion of restricted maximum likelihood. We demonstrate the performance of these new metrics for the estimation of Bayesian networks using simulated data with  known complex mean structures.  We then present the  analysis of expression levels of grape berry genes adjusting for exogenous variables believed to affect the expression levels of the genes. Demonstrable biological effects can be inferred from the estimated conditional independence  relationships and correlations amongst the grape-berry genes.
\end{abstract} 

Bayesian network; complex mean structure; exogenous variable; grape-berry gene expression; regulatory network; score-based metric; variance components. 

\section{Introduction}\label{IntroSection}

The inner workings of a cell are very complex, with many interacting components. Determining how the genes within a cell interact with each other is an important, but difficult, field of research, often requiring the application  of advanced statistical methods.  Systems of these gene interactions are known as genetic regulatory networks, and the extent to which such networks may be inferred from observational gene expression data remains largely undetermined.  To explore this question carefully and quantitatively, high-dimensional multivariate models, including Bayesian networks, need to be considered. 
The use of Bayesian networks for the modelling of genetic regulatory networks has been discussed by several authors: see for example \cite{ HiddeJong, FriedmanScience, Friedman,  Markowetz}. Their popularity lies in the provision of a flexible framework for the estimation of conditional dependence relationships, thereby providing a means to estimate a covariance matrix given a high-dimensional sample when maximum likelihood methods are unavailable, \cite{Dykstra}. Estimation of such structures allows insight into how the expression levels of large groups of genes are related to one another, which, in turn, should help shed light on genetic regulatory networks involving the genes. 

For the most part, it has been assumed that the data used to estimate the networks are independent and identically distributed. In the present paper, we consider the important case where the assumption of independent and identically distributed samples is not satisfied, and propose new methods to allow for the estimation of effects of interest given such complexity.  Our theoretical development has been motivated by an observational time course microarray study, involving  expression levels of grape-berry genes observed over time and known to be associated with changes in temperature.  The grapes  were sampled from three vineyards in different regions of  southern Australia and data on the ambient temperatures during the times leading up to the picking of each sample of grapes was also measured. We want to investigate the conditional dependence structure of the genes, adjusting for the exogenous effects of temperature and vineyards,  and we aim to do this through the estimation of a Bayesian network.  If the effect of temperature is unaccounted for in the estimation of a Bayesian network for these genes, because of their common relationship with temperature, many pairs of genes will exhibit strong correlations.  Unless the gross effects of vineyard and temperature are removed, one cannot hope to detect more subtle associations between genes.

There are many methods available for the estimation of Bayesian networks given microarray  and other high-dimensional datasets, and these may be divided into two broad categories, namely,  score-based and constraint-based methods, \cite{SGS}. Score-based methods attempt to maximize some score metric associated with the estimated Bayesian network, whilst constraint-based methods estimate conditional independence relationships directly from the data, and combine these to form a Bayesian network.  Constraint-based methods test for conditional independence relationships, so the networks obtained through their application can be quite sensitive to Type I and Type  II errors, particularly when the sample sizes are small. Score-based methods on the other hand  are not as sensitive to small sample sizes, and instead of finding the best local structure for each node, find the best global structure given the data, often resulting in more parsimonious models. Given that gene expression data sets tend to be high-dimensional with the attendant \lq small $n$, large $p$' problem, we approach the problem of Bayesian network estimation from a score-based perspective, and extend these to include exogenous variables and dependent data.

{The outline of the paper is as follows. In Section \ref{ReviewSection}, Bayesian networks and score metrics are briefly reviewed, and our two new score metrics for datasets with complex mean structure and random effects are presented. The new score metrics are  used to estimate Bayesian networks for simulated datasets with a known complex mean structure in Section \ref{SimulatedSection}, and  then applied to the analysis of the grape-berry gene expression data in Section \ref{GrapeDataAnalysis}. In Section \ref{ConclusionSection}, we present a brief summary of our overall findings.}

\section{Bayesian networks and Score Metrics}\label{ReviewSection}

\subsection{BGe, the basic Bayesian score metric}
Consider a random vector ${X} = ( X_1,  \ldots, X_p)^T$. A Bayesian network $B$ for ${X}$ consists of two components: a directed acyclic graph ${G} = (V,E)$ with $V = \left\{ X_1,  \ldots, X_p\right\}$, often written as $V = \left \{1, \ldots, p \right\}$, and assumed conditional distributions $f\left( x_i \mid {x}_{P_i}, \theta_i\right), i = 1, \ldots, p$. The set $P_i$ is the set of parents of $X_i$ in ${G}$ and $\Theta = \left\{\theta_1, \ldots, \theta_p\right\}$ is the set of parameters associated with the conditional distributions. The graph and conditional distributions then specify a joint distribution for ${X}$:
\begin{equation}
f\left({x} \mid G, \Theta \right) = \prod_{i=1}^p f\left( x_i \mid {x}_{P_i}, \theta_i\right). \label{BNeq}
\end{equation}
Bayesian networks encode information about the conditional independence relationships between the variables in ${X}$. The directed Markov properties, as described in \cite{Lauritzen}, for example, allow conditional independence statements about ${X}$ to be read from the graph ${G}$. Additionally, when the available data set is high-dimensional, and maximum likelihood estimation of the covariance matrix of $X$ is unavailable, estimation of a Bayesian network allows the estimation of the covariance matrix.

Here we consider a vector of pre-processed, normalized expression levels for $p$ genes, and suppose that ${X}\sim N_p ({0}, \Sigma)$, where $\Sigma$ is unknown. We consider a data set $d = \left\{{x}_1, \ldots, {x}_p \right\}$, where ${x}_i = \left( x_{i1}, \ldots, x_{in}\right)^T$ is the vector containing the $n$ samples of the expression levels of gene $i$. The estimation of a Bayesian network for ${X}$ given this data set $d$ consists of learning the structure of a directed acyclic graph encoding the conditional independence relationships between the variables, and the estimation of the parameters $\Theta$.

As described in the Introduction, a  score-based approach to learning the structure of the graph encoding the conditional dependence relationships of $X$ is taken. After deciding on a score metric, we want to find a graph that maximises that score metric, and an obvious choice is the likelihood function of directed acyclic graphs given the data. The maximum likelihood score turns out not to be a good idea, however, as it inevitably assigns the complete graph, encoding no conditional independence relationships, the maximum score. For more detail, readers are directed to Section 18.3 of \cite{KollerFriedman}.

To avoid the problems with overfitting associated with the maximum likelihood score, the Bayesian score was developed. Following \cite{LearningGaussianNetworks} among others, the Bayesian score metric for the estimation of a directed acyclic graph ${G}$ given some data set $d$ is proportional to the posterior probability of that graph:
\begin{equation}
S({G} \mid d) = p({G})f(d \mid {G}) = p(G) \int_{\mathbb{R}^{np}} f(d \mid {G}, \Theta) f(\Theta \mid {G}) d \Theta. \label{scoremetric}
\end{equation}
Here the focus is on the second component of this score, the marginal model likelihood of the data given the graph ${G}$, where the density of $d$ given $G$ and $\Theta$ is assumed to be an $np$-dimensional normal density, with mean vector $0$ and covariance matrix $\Sigma \otimes I_n$. As per Equation \eqref{BNeq}, this joint density may be decomposed into a product of $p$ conditional densities. When the data set $d$ consists of independent and identically normally distributed samples, $\theta_i = \left\{{\gamma}_i, \psi_i \right\}$, and
\[
{x}_i \mid {x}_{P_i},{\gamma}_i, \psi_i\sim N_n \left({x}_{P_i} {\gamma}_i, \psi_i I_n\right).
\]
Given normal-inverse gamma priors for each $\theta_i$, or an equivalent inverse Wishart prior on $\Theta$, the score metric of Equation \eqref{scoremetric} can be written as the product of the prior density on the space of directed acyclic graphs, $p(G)$, and $p$ multivariate $t$ densities.
This score metric, only appropriate in the case of independent and identically distributed samples, is known as the BGe metric: \lq\lq Bayesian metric for Gaussian networks with score equivalence\rq\rq.

\subsection{BGeCM, the score metric for data sets with complex mean structure}

We now consider the case of a more complex data set $d$, that does not consist of independent and identically distributed samples, such as the grape-berry gene data set described in Section \ref{IntroSection}. As explained there, contained within that data set is information about exogenous variables thought to affect the expression levels of the genes under study. 
Given such a data set, we now express the model for the vector of expression levels of gene $i$ as
 \begin{equation}
{x}_i \mid {x}_{P_i}, {\gamma}_i, \psi_i, {b}_i, \phi_i \sim N_n \left({x}_{P_i}{\gamma}_i + Q{b}_i, \psi_i I\right), \label{sitemodel}
\end{equation}
where ${b}_i$ is the $m$-vector of the effects of the $m$ exogenous variables on gene $i$, $\phi_i$ are the parameters associated with the (as yet to be selected) prior distribution for ${b}_i$, and $Q$ is the $n \times m$ matrix containing the data associated with the $m$ exogenous variables. It can be seen that in this specification, we retain linear dependence upon expression levels of parent genes, but now, in addition to that dependence, more complex sampling schemes and the influence of exogenous variables are accounted for through the linear dependence of expression levels upon   ${b}_i$.

Including exogenous variables in the estimation of a Bayesian network is important in order to obtain an unbiased estimate of the conditional dependence relationships between the genes of interest. For example, the expression levels of two genes may both be dependent upon changes in an exogenous variable, but conditionally independent of each other. If dependence upon exogenous variables is not accounted for, an edge between these two genes is likely to be present in an estimated graph. By accounting for the effects of exogenous variables, we can have more confidence that the conditional dependence relationships obtained represent actual dependence relationships, and are not due to common relationships with exogenous variables.

As can be seen by the definition of the Bayesian score metric given by Equation \eqref{scoremetric}, a joint prior distribution for ${\gamma}_i, \psi_i,{b}_i$ and $\phi_i$ is required for the calculation of a score metric. Care is required in the specification of this prior distribution, since if priors are not properly selected, a score metric that gives different scores to directed acyclic graphs encoding equivalent conditional independence restrictions will be induced. A score metric that does not discriminate between equivalent directed acyclic graphs is called an equivalent score metric. Discrimination between equivalent graphs is tantamount to assigning causal meaning to the directed edges of ${G}$, and since the emphasis here is on the estimation of graphs given observational data, the assignation of causal meaning to the estimated relationships is not appropriate.

Extension of the results in \cite{GeigerHeckerman} to our model indicates that the joint prior distribution for ${\gamma}_i, \psi_i,{b}_i$ given $\phi_i$ must have a normal-inverse gamma form, and that the effects of the exogenous variables on one gene must be \emph{a priori} independent of the effects upon another gene, for the induced score metric to satisfy equivalence. The following system of priors are used:
\begin{align}
{\gamma}_i \mid \psi_i \sim N_{|P_i|} \left({0}, \tau^{-1}\psi_i I_{| P_i|} \right),&\quad 
\psi_i^{-1}  \sim Ga \left( \frac{\delta +|P_i|}{2}, \frac{\tau}{2}\right),\notag \\ 
{b}_i \mid \phi_i &\sim N_m \left({0}, \phi_i I\right).  \notag
\end{align}
There are several possible assumptions about the form of the prior distribution of the variance of the random effects for gene $i$, $f(\phi_i)$. Among other choices, the variance of the random effects could be assumed known, a uniform prior could be placed on $\sqrt{\phi_i}$, or an inverse gamma prior could be placed on $\phi_i$. However, by an extension of the results in  \cite{GeigerHeckerman},  when $\phi_i \neq \upsilon^{-1}\psi_i$, any choice of prior distribution on $\phi_i$ will result in a marginal model likelihood without a closed form, requiring numerical integration to compute and slowing down computations.

A simulation study in \cite{Kasza} showed that the learnt network structure is quite robust to the misspecification of the prior density for $\phi_i$, provided the magnitude of $\phi_i$ is correctly specified. Hence, for computational simplicity, the following prior for the variance of the effects of exogenous variables is used:
\[
{b}_i\mid \phi_i \sim N_m \left( 0, \upsilon^{-1}\psi_i I  \right),
\]
 where $\upsilon$ is some positive parameter that is constant from gene to gene. If $\upsilon = \tau$, then ${b}_i$ and ${\gamma}_i$ are independent and identically distributed. Taking $\upsilon > \tau$ implies that the ${b}_i$ are less variable than the ${\gamma}_i$, while $\upsilon < \tau$ implies that the ${b}_i$ are more variable than the ${\gamma}_i$. If $\upsilon$ is taken to be very large, this is equivalent to assuming that the effects of exogenous variables do not contribute much to the overall variability of the expression levels of genes.

Although it will be application dependent,  it may be that the assumption  the variance of the exogenous variables is related to the variance of the regression parameters in the same way for each gene is not be valid. In this situation,  a separate $\upsilon_i$ could be specified for each gene, but such specification would require information that is most probably unavailable. Alternatively, a hyperprior distribution could be placed upon the $\upsilon_i$. However, any choice of such a distribution would lead to a score metric without an exact form, again requiring numerical integration to compute. 

When ${b}_i \sim N_m ({0}, \upsilon^{-1} \psi_i I)$, the marginal model likelihood for a particular random variable given its parents in the graph $G$, can be shown to be
\[
{x}_i \mid {x}_{P_i}\sim t_{\delta + | P_i|}\left({0}, \Sigma_{{x}_i \mid {x}_{P_i}} \right), 
\]
with
\begin{align}
\Sigma_{{x}_i \mid {x}_{P_i}} &= \frac{\tau}{\delta + |P_i| }\left\{ J- J{x}_{P_i}\left( \tau I + {x}_{P_i}^TJ {x}_{P_i}\right)^{-1} {x}_{P_i}^TJ\right\}^{-1},\notag \\
J &=  I -  Q \left( \upsilon I + Q^TQ \right)^{-1}Q^T. \label{Jdef}
\end{align}
We call the resultant score metric the BGeCM metric: \lq \lq Bayesian metric for Gaussian networks having score equivalence for data sets with a complex mean structure\rq\rq.

Posterior distributions of the parameters ${\gamma}_i $, $\psi_i$ and ${b}_i$ allow a detailed analysis of the relationships between random effects and the expression levels of the genes of interest. The posterior distributions of ${\gamma}_i$ given $\psi_i$, ${b}_i$ given $\psi_i$ and $\psi_i$ are given by
\[
{\gamma}_i\mid {x}_i, \psi_i, {x}_{P_i} \sim N_{| P_i|}\left( \left(\tau I + {x}_{P_i}^T J {x}_{P_i} \right)^{-1}{x}_{P_i}^T J {x}_i,   \psi_i \left( \tau I + {x}_{P_i}^T J {x}_{P_i} \right)^{-1}\right), 
\]
where $J$ is as given in Equation (\ref{Jdef}). Further,
\begin{align}
{b}_i \mid {x}_i, \psi_i, {x}_{P_i} &\sim N_m \left(\left(\upsilon I + Q^T J^\ast Q \right)^{-1}Q^T J^\ast {x}_i,   \psi_i \left( \upsilon I + Q^T J^\ast Q \right)^{-1} \right), \notag \\
J^\ast &= I - {x}_{P_i}\left( \tau I + {x}_{P_i}^T{x}_{P_i}\right)^{-1}{x}_{P_i}^T, \notag
\end{align}
and
\begin{align}
\psi_i \mid {x}_i, {x}_{P_i} &\sim \text{ Inv Gamma}\left(\frac{n + |P_i| +\delta}{2}, \beta_{\psi_i} \right), \notag \\
\beta_{\psi_i}&= \frac{\tau }{2}+ \frac{1}{2}{x}_i^T \left\{J - J {x}_{P_i}\left(\tau I + {x}_{P_i}^T J {x}_{P_i} \right)^{-1}{x}_{P_i}^T J\right\} {x}_i. \notag
\end{align}

Note that instead of using a score metric as developed above to allow for the inclusion of exogenous variables in the model, an extended directed acyclic graph could be learnt, where exogenous variables are included as vertices in the graph. There are however, a couple of difficulties presented by such an approach. The first is that if the exogenous variables are discrete, methods for Bayesian networks on both continuous and discrete variables are required. Additionally, many algorithms for learning directed acyclic graphs incorporate sparsity constraints, and if it is believed that many of the genes are affected by these exogenous variables, these sparsity constraints will require modification.

\subsection{Removal of random effects through analysis of residuals}

In the derivation of the BGeCM score metric, it was assumed that the effects of exogenous variables on gene expression were of intrinsic interest. However, in many situations, the effects of exogenous variables can be thought of as nuisance variables, complicating the estimation of Bayesian networks for the given gene expression levels. It may be desirable to ignore the possible influences of such effects upon gene expression levels, and on the relationships between genes. Of course, simply ignoring such effects is not recommended. Instead, we develop a non-parametric approach that adjusts for the effects of exogenous variables, without making assumptions about the form of their distributions.
This approach, instead of directly using the gene expression data, is based upon the use of linear combinations of residuals left over after the data is regressed upon the effects of the exogenous variables. We call this the \lq\lq residual approach\rq\rq, and it is inspired by the restricted maximum likelihood procedure used in inference for mixed linear models; see for example Section 12.2 of \cite{Davison}, or Speed \cite{Speed:1997aa}, which provides a good overview of REML. 

The utility of the residual approach is that it makes no assumptions about the distributional form of the random effects of interest.  Since no such assumptions are made, the approach is correct no matter what the true distribution of the random effects may be. Hence, in situations when the assumption that ${b}_i\mid \phi_i \sim N_m \left({0}, \upsilon^{-1}\psi_i I  \right)$ is not satisfied, the residual approach provides a useful alternative to the BGeCM score metric, and, as we demonstrate below, is considerably easier to implement.

We consider an $(n-m)\times 1$ random variable ${y}_i = P^T {x}_i$, where $P$ is an $n \times (n-m)$ matrix such that
\[
P^TQ =  0, \quad
P^TP = I_{n-m},\quad
PP^T =  I_n - Q(Q^TQ)^{-1}Q^T. \label{Pconds}
\]
Hence, 
\[
{y}_i \mid {\gamma}_i, \psi_i, {y}_{P_i} \sim N_{n-m} ({y}_{P_i} {\gamma}_i , \psi_i I),   \]
and the score metric associated with this set of marginal model likelihoods is invariant to the choice of $P$.
Implementation of the residual approach to the estimation of Bayesian networks is therefore simple: after selection of an appropriate matrix $P$ and computation of ${y}_i = P^T {x}_i$ for $i=1, \ldots, n$, the BGe score metric may be applied to this reduced data set in conjunction with the score-based method of choice.

A  drawback of the residual approach is that posterior estimates of the random effects ${b}_i$ are not admitted. However, any potential loss of information about the underlying covariance matrix when the residual approach is used, compared to the \lq full' BGeCM score metric, has been investigated in \cite{KaszaSolomon}, and  found to be typically small.

\section{Numerical study of BGeCM and the residual score metrics}\label{UseSection}

\subsection{Implementation of BGeCM and the residual score metrics}

In this section, the necessity of score metrics that take account of complex mean structure are demonstrated through the application of the residual approach and the BGeCM score metric to simulated and real data sets.

First, a note on implementation. The BGeCM score metric and the residual approach may be incorporated into any score-based algorithm for the estimation of Bayesian networks, without the need for any additional programming. In the case of the residual approach, all that is required is the calculation of the matrix $P$, satisfying the conditions in Equation (\ref{Pconds}). Then, instead of inputting $d$ into the algorithm of choice, the augmented data set $P^Td$ is input. Similarly, when the BGeCM score metric is used, an augmented data set $L^Td$ will be the input into the algorithm, where $L^T$ is a matrix such that $J = LL^T$, where $J$ is as given in Equation (\ref{Jdef}).

Here we apply the residual approach and BGeCM score metrics in conjunction with the high-dimensional Bayesian covariance selection algorithm, \cite{SAMSI}, a score-based method for the estimation of Bayesian networks. This algorithm works by constructing and combining regression models for each $X_i$.

 \subsection{Simulated data sets}\label{SimulatedSection}

\emph{Example 1:} In this first example, 10 data sets were generated according to the following system of linear recursive equations: 
\[
X_{ijk} = b_{ij}+ \epsilon_{ijk}, \quad \epsilon_{i}\sim  N(0, \psi_i) \quad  (i=1, \ldots, 100;\ j =1,2;\ k=1,\ldots, 50).
\]
The values of $\psi_i$ were obtained by sampling from an Inverse Gamma$(1,1/2)$ distribution, and are constant for each of the samples generated. Similarly, ${b}_i = \left(b_{i1}, b_{i2} \right)^T$, $i = 1, \ldots, 100$, are fixed across data sets, obtained by sampling from 
\[
b_{ij}\sim  N(0, \psi_i) \quad  (i =1, \ldots, 100; \ j =1,2),\]
corresponding to $\upsilon =1$. The non-zero mean structure of this example corresponds to two groups, and the true underlying directed acyclic graph is the empty graph.

\emph{Example 2:} The system of linear recursive equations governing this example is 
\[
X_{ik} = q_{1k}b_{i1}+q_{2k}b_{i2}+ q_{3k}b_{i3}+ \epsilon_{ik}   \quad (i =1, \ldots, 18), \]
\[
X_{19,k} = q_{1k}b_{19,1}+q_{2k}b_{19,2}+ q_{3k}b_{19,3}+ \gamma_{19,1}X_{1k}+ \gamma_{19,2}X_{2k}+\epsilon_{19,k}, \]
\[
X_{20,k} = q_{1k}b_{20,1}+q_{2k}b_{20,2}+ q_{3k}b_{20,3}+ \gamma_{20,19}X_{19,k}+\epsilon_{20,k}, \]
\[
\epsilon_{ik}\sim  N(0, \psi_i) \quad (i =1, \ldots, 20;\ k =1,\ldots, 10). \]  
Ten data sets were generated according to this system of equations, and the parameters $\psi_i \ (i =1, \ldots,20)$, ${\gamma}_{19} = (\gamma_{19,1}, \gamma_{19,2})^T$ and $\gamma_{20,19}$ were assumed constant across these data sets. The values of these parameters were obtained by sampling from the following distributions:
\[
\psi_i \sim  \text{Inv Gamma}\left(\frac{2+|P_i|}{2}, \frac{1}{2}\right), \quad |P_i | = 0 \quad (i =1, \ldots, 18), \quad | P_{19}|=2, \quad | P_{20}|=1, \]
\[
{\gamma}_{19}  \sim N_2 \left( {0}, \psi_{19}I_2\right), \gamma_{20,19} \sim N(0, \psi_{20}). \] 
Similarly, the random effects ${b}_i = \left( b_{i1}, b_{i2}, b_{i3}\right)^T \ (i =1, \ldots, 20)$, were constant across the 10 data sets generated, obtained by sampling from 
\[
b_{ij}\sim  N(0, \psi_i) \quad (i =1, \ldots, 20;\ j =1,2, 3), 
\]
again corresponding to $\upsilon =1$.

The true model for each variable may be written as
\[
{x}_i \mid  \psi_i , {b}_i \sim N_{10}\left(Q{b}_i, \psi_i I_{10} \right) \quad (i =1, \ldots, 18), \]
\[
{x}_{19} \mid{\gamma}_{19} ,  \psi_{19} , {b}_{19} \sim N_{10}\left({x}_{P_{19}}{\gamma}_{19} + Q{b}_{19}, \psi_{19} I_{10} \right), \]
\[
{x}_{20} \mid{\gamma}_{20} ,  \psi_{20} , {b}_{20} \sim N_{10}\left({x}_{P_{20}}\gamma_{20} + Q{b}_{20}, \psi_{20} I_{10} \right), \]
where 
\[
{x}_i =\left( \begin{array}{c} x_{i1} \\  \vdots \\ x_{i10}  \end{array}\right), \]
\[
{x}_{P_{19}} = \left( \begin{array}{c c} {x}_1, {x}_2 \end {array}\right), \quad {x}_{P_{20}} =  {x}_{19} \]
and 
\[
Q = \left( \begin{array}{ccc}
q_{11}& q_{21}& q_{31}\\
\vdots & \vdots & \vdots \\
q_{1,10}& q_{2,10}& q_{3,10}
\end{array}\right)
 =\left( \begin{array}{rrr}
-1\cdot 32&  0\cdot 83& -1\cdot 74 \\
   0\cdot 22& -1\cdot 37 & 0\cdot 55\\
   0\cdot 37 & 0\cdot 61 &  0\cdot 60 \\
    -1\cdot 53 &  1\cdot 52 &  0\cdot 82\\
     -0\cdot 73 &-0\cdot 01 & 0\cdot 93\\
0\cdot 92  & 0\cdot 87&  -0\cdot 09 \\
 1\cdot 02 & -0\cdot 44& -0\cdot 04\\
 0\cdot 27 & -0\cdot 59& 0\cdot 11 \\
  -0\cdot 64 & 0\cdot 20 & -0\cdot 21 \\
  -0\cdot 15 & 0\cdot 48 & -0\cdot 12
 \end{array}\right). 
\]
In this case, elements of the $Q$ matrix consist of random samples from the standard normal distribution, treated as known constants in the analysis, and the true underlying graph has three edges.

Bayesian networks were estimated for each of the data sets generated according to Examples 1 and 2, using the BGe and BGeCM score metrics and the residual approach in conjunction with the High-dimensional Bayesian Covariance Selection algorithm. After assessing the performance of  the BGeCM score metric under ideal conditions, we assess the sensitivity of this metric to the misspecification of $\upsilon$.

For each of these analyses, the number of spurious and correct edges in the highest-scoring network found by the algorithm was recorded. The results are summarized in Tables \ref{ResultsTable} and \ref{Table2}. Table \ref{ResultsTable} gives the mean and standard deviation of the number of spurious and correct edges in the highest-scoring Bayesian networks found when BGe,  BGeCM and the residual approach, with $\upsilon$ set at the correct value, that is $\upsilon=1$, are used to estimate the true graph encoding the conditional independence relationships. Table \ref{Table2} gives the numbers of correct and spurious edges when BGeCM is used given a range of values of $\upsilon$.

  \begin{table}
  \def~{\hphantom{0}}
\caption{Mean and standard deviation of the number of spurious and correct edges in the highest-scoring Bayesian networks obtained when the score metrics are applied to data sets simulated according to Examples 1 and 2\label{ResultsTable}}{
\begin{tabular}{  c | c | c c }
& {Example 1}& \multicolumn{2}{c}{Example 2} \\  
Score Metric & Spurious Edges  &  Correct Edges & Spurious Edges  \\ \hline
BGe & 117$\cdot$ 2(5$\cdot$ 25)& 1$\cdot$ 8(0$\cdot$ 63) &2$\cdot$ 4(1$\cdot$ 17) \\
BGeCM & ~~ 1$\cdot$ 0(0$\cdot$ 94) &2$\cdot$ 2(0$\cdot$ 92) & 0$\cdot$ 4(0$\cdot$ 70)\\
Residual Approach & ~~1$\cdot$ 7(1$\cdot$ 16)  &  1$\cdot$ 1(0$\cdot$ 32) & 0$\cdot$ 0(0$\cdot$ 00)
\end{tabular}}
\end{table}

\begin{table}
  \def~{\hphantom{0}}
\caption{Mean and standard deviation of the number of spurious and correct edges in the highest-scoring graphs found through the application of BGeCM for varying values of ${\upsilon}$, (standard deviation in brackets).\label{Table2}}{
\begin{tabular}{l | c c c c c  c }
 & \multicolumn{6}{c}{${\upsilon}$   }  \\  
 {Example}   &  0$\cdot$0001 & 0$\cdot$001  &  0$\cdot$01  & 0$\cdot$1 & $ 1 $ & $2 $\\ \hline
{1 - Spurious} & 1 $\cdot$4(0$\cdot$97) & 1$\cdot$4(0$\cdot$97) & 1$\cdot$4(0$\cdot$97) & 1$\cdot$3(0$\cdot$82) &1$\cdot$0(0$\cdot$82) & 1$\cdot$4(1$\cdot$07) \\
2 - Spurious & 0 $\cdot$9(1$\cdot$10)&0$\cdot$8(0$\cdot$92)&1$\cdot$0(0$\cdot$82)&0$\cdot$9(0$\cdot$99)&0$\cdot$4(0$\cdot$70)&0$\cdot$6(0$\cdot$70) \\
2 -  Correct  & 1 $\cdot$9(0$\cdot$57)&2$\cdot$2(0$\cdot$63)&2$\cdot$2(0$\cdot$63)&1$\cdot$9(0$\cdot$74)&2$\cdot$2(0$\cdot$92)&2$\cdot$1(0$\cdot$74)  \\
 & \multicolumn{6}{c}{${\upsilon}$   }  \\  
{Example} & $ 5$ &  $10 $ & $ 20 $ & $50$ & $100$ & 1000 \\  \hline
{1 - Spurious}  & 3 $\cdot$8(2$\cdot$39) & 15$\cdot$8(3$\cdot$46) & 45$\cdot$6(5$\cdot$76) & 84$\cdot$6(4$\cdot$17) & 101$\cdot$7(4$\cdot$42) &  119$\cdot$4(2$\cdot$88) \\
2 - Spurious & 0 $\cdot$4(0$\cdot$52)&~1$\cdot$1(0$\cdot$88)&~1$\cdot$3(1$\cdot$16)&~2$\cdot$1(0$\cdot$88)&~~2$\cdot$3(1$\cdot$25)&~~2$\cdot$2(1$\cdot$48)\\
2 - Correct &2 $\cdot$5(0$\cdot$53) & ~2$\cdot$1(0$\cdot$74) & ~2$\cdot$2(0$\cdot$79)&~2$\cdot$0(0$\cdot$82) & ~~1$\cdot$8(0$\cdot$79) &  ~~2$\cdot$0(0$\cdot$82) 
 \end{tabular}}
\end{table}

Comparing the results obtained when the BGe score metric is used to analyse the simulated data sets demonstrates the utility of both  the BGeCM score metric and the residual approach. The  two new score metrics result in the estimation of structure which is much closer to the true structure. In addition, Table \ref{Table2} shows that the results obtained from the BGeCM score metric are quite robust to the misspecification of $\upsilon$, producing accurate results when the value of $\upsilon$ selected departs as much as one or two orders of magnitude from its true value. As $\upsilon$ gets larger, the highest scoring graphs obtained become more and more similar to those obtained when the BGe metric is used. This is a result of the fact that as $\upsilon$ approaches $\infty$, the limit of the BGeCM metric is the BGe metric, \cite{KaszaSolomon}.

 \section{Analysis of the grape-berry microarray data}\label{GrapeDataAnalysis}
 
The data analysed here  consisted of $50$ samples of gene expression levels for $26$ grape genes measured over a four-week period. The gene expression levels were derived from grape-berry tissue samples grown in  three different vineyards in three different wine-growing regions of southern Australia.  Twenty samples were taken  from a vineyard in Clare, $20$ from the Wingara Vineyard in Mildura and $10$ from a vineyard in Willunga.  Table \ref{HeatShockGenesTable} provides the reference numbers for the $26$ grape genes, together with a brief summary of their functions. All  the genes in Table \ref{HeatShockGenesTable}  are known to code for heat shock proteins (HSPs), \cite{HSPs}, which are responsible for protecting the grapes against heat-induced stress.   In addition to data on the gene expression levels, temperature in degrees celcius was recorded during the time leading up to the picking of the grapes.

These data are part of a larger dataset on grape-berry tissue samples measured between 2003 and 2005. At each vineyard, grapes were sampled roughly weekly over the period of  development of the berries, {\it i.e.},  from the time buds formed on the vines, to the time when the grapes were ripe.   In general, grape berries follow a double sigmoidal pattern of growth that consists of two distinct growth phases, with a lag period between these phases (see \cite{CoombeGrape, DaviesGrape}). The second stage of grape-berry growth commences upon the occurrence of veraison, when the grape berries start to change colour. Robinson and Davies, \cite{DaviesGrape}, suggest that at veraison and during ripening, there are many changes in the expression levels of many different genes in grape berries. The observed developmental  time period  from bud formation to grape ripeness  differed between vineyards in the present study. The shorter four-week time period we analysed occurred after fruit set, but well before veraison for all three vineyards.   We  restricted attention to samples corresponding to the third to seventh sampling weeks at each vineyard because the relationships between genes are thought to be more stable during this period, and the modelling assumption of 
identically distributed samples is therefore more likely to be valid.

mRNA expression levels for each of the grape tissue samples was measured using Affymetrix Vitis vinifera oligonucleotide arrays. Background subtraction and normalisation was carried out using robust microarray analysis (RMA), as described in Irizarry \emph{et al}, \cite{Irizarry}. Note that all samples were processed at the same laboratory.

Understanding the stress tolerance mechanisms of plants is important, and the heat shock protein network, as discussed by  \cite{KotakHSP} and  \cite{HSPs}, is very complex.  The heat shock protein network of plants is believed to consist of interactions between small Hsps, Hsp60, Hsp70, Hsp90  and Hsp100,  \cite{HSPs}.  Precisely how Hsps interact with one another and how they protect against heat stress is not yet completely understood, and here we seek to gain some insight into the heat shock protein network by examining the conditional dependence structure of the genes given in Table \ref{HeatShockGenesTable}.

 \begin{table}[h]
\begin{center}
\caption[Grape heat shock genes.]{The grape heat shock genes. The first column gives the gene reference numbers used in this study, the second column gives the Affymetrix reference numbers, and the third column gives the National Center for Biotechnology Information reference numbers. The fourth column provides a short description of the function of the genes. Note that HSP stands for heat shock protein.}
\label{HeatShockGenesTable}
\footnotesize{
\begin{tabular}{c r  l l } 
Ref $\#$ & Affymetrix $\#$&NCBI $\#$	&Protein Identity	 \\ \hline

1&1616246$\_$at&	Vvi.9142&	Heat shock protein 70, ATP binding\\ \hline
2&1607002$\_$at&	Vvi.4801&	Heat shock protein 70, ATP binding, \\
 & & & luminal binding protein, glucose regulated \\ \hline
		
3&1610684$\_$at&	Vvi.2869&	chloroplast HSP 70-1, ATP binding	\\ \hline

4&1611740$\_$at&	Vvi.295	&unknown\\ \hline 

5& 1620985$\_$at&	Vvi.4530&	HSP21 chloroplast\\ \hline  
 
6& 1616995$\_$at&	Vvi.23518&	\multirow{3}{*}{Ubiquitin conjugating enzyme 4e}	\\ 
7&1614132$\_$at&	Vvi.863&	\\
8&1618265$\_$at&	Vvi.15427&	\\ \hline

9&1608052$\_$s$\_$at&	Vvi.9085&\multirow{3}{*}{HSP81(early response to dehydration)}\\ 
10&1618009$\_$at&	Vvi.9085& \\
11&1619931$\_$s$\_$at&	Vvi.7394&\\ \hline

12&1608701$\_$at&	Vvi.2083&	10 kDa chaperonin \\ \hline

13&1608164$\_$at&	Vvi.6787&\multirow{5}{*}{Cytosolic class II 17.6 HSP}\\
14&1611052$\_$at&	Vvi.6787&\\
15&1611192$\_$at&	Vvi.6787&\\
16&1610032$\_$at&	Vvi.6787&	\\
17&1614330$\_$at&	Vvi.6787&\\ \hline 

18&1620956$\_$at&	Vvi.3921&\multirow{5}{*}{17.6 kDa class I small HSP}\\
19&1616538$\_$at&	Vvi.7869&	\\
20&1609554$\_$at&	Vvi.7044&	\\
21&1620960$\_$a$\_$at&	Vvi.7044&	\\
22&1621652$\_$at&	Vvi.4464&	\\ \hline
 
23&1622165$\_$at&	Vvi.6156&\multirow{2}{*}{17.4kDa class I small HSP}	\\
24&1612385$\_$at&	Vvi.4422&	\\ \hline

25&1622628$\_$at&	Vvi.5040&	17.4kDa class III small HSP\\ \hline

26&1610700$\_$at&	Vvi.2537&	23.6K mitochondrial small HSP\\ \hline
\end{tabular}}
\end{center}
\end{table}

Given the known functions of the genes considered in this study and the climatic and geographic disparities between the regions where the grape berries were sampled, it would incorrect to ignore the effects of vineyard and temperature in the estimation of a Bayesian network for the grape genes. The essential point is that if the expression levels of these genes are strongly influenced by these exogenous variables, then accounting for variation due to such variables  in the estimation of a Bayesian network should result in a network that more accurately encodes the dependence structure of the genes. A further important point is that given the grape gene expression levels analysed are observational data, causal interpretations should not be applied to the directed edges present in any network estimated given the data. Hence, moralized versions of directed acyclic graphs are used to summarize conditional independence relationships of the grape genes.

To begin, the  initial (null) model omitted the effects of vineyard and temperature on the expression levels of the genes. That is, if ${x}_i$ is the 50-vector of the expression levels for grape gene $i$, it is assumed that
\begin{align}
&x_i \mid {x}_{P_i}, {\gamma}_i, \psi_i \sim N_{50}\left({x}_{P_i} {\gamma}_i, \psi_i I_{50}\right), \notag \\
& {\gamma}_i \mid \psi_i \sim N_{|P_i|}({0}, \tau^{-1}\psi_i I_{|P_i|}), \quad \psi_i ^{-1} \sim Ga \left( \frac{\delta +| P_i|}{2}, \frac{\tau}{2} \right), \notag \end{align}
where $ {x}_{P_i}$ is a $50 \times |P_i|$ matrix. The columns of this matrix consist of the expression levels of the grape genes in the dataset that the expression level of gene $i$ is dependent upon, ${\gamma}_i = \left( \gamma_{ij}\right)_{j \in P_i}$ and $\gamma_{ij}$ is the effect of the expression level of gene $j$ on the expression level of gene $i$. Following the analysis of Affymetrix gene expression data in \cite{SAMSI}, $\tau = 1$ and $\delta = 2$.

\begin{figure}
\vspace{-1cm}
\begin{center}
\includegraphics[width=60mm]{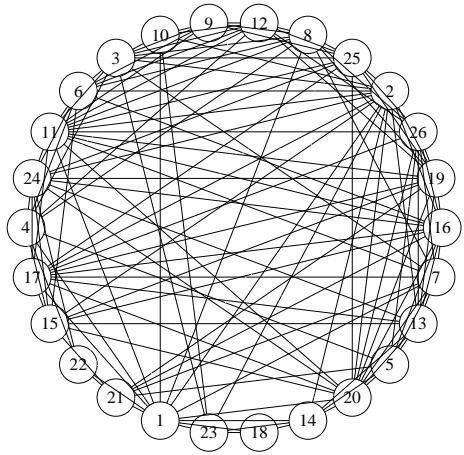}\\
(a)  \\ \vspace{0.25cm}
\includegraphics[width=60mm]{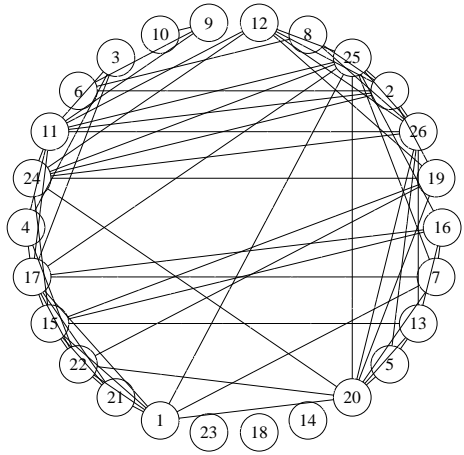} \\
(b)\\\vspace{0.25cm} 
\includegraphics[width=60mm]{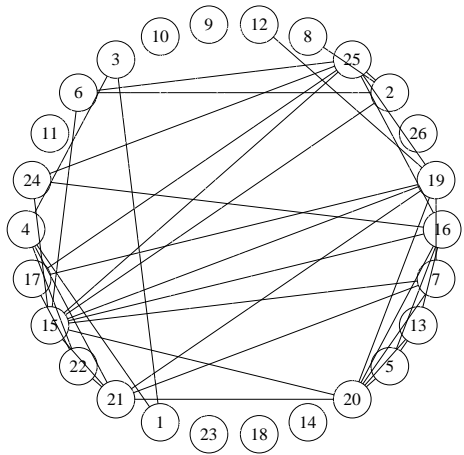} \\
\mbox{(c)} 
\end{center}
\caption{The moral versions of the highest-scoring graphs obtained for the grape genes when (a) the effects of temperature and vineyard are ignored, and when the residual approach is used to include (b) vineyard effects, and (c) vineyard effects and main temperature effects.} \label{GrapeGraphs}

\end{figure}

The highest scoring Bayesian network found through the application of the high-dimensional Bayesian covariance selection algorithm to the full dataset (ignoring the exogenous variables of vineyard and temperature) has $55$ edges, and the moralized version has $130$ edges. The moralized version is shown in Figure \ref{GrapeGraphs}(a).  

Next, graphs were  estimated separately for each of the three  vineyards. The highest-scoring directed acyclic graphs obtained for the Clare, Wingara and Willunga vineyards had, respectively, $22, 23$ and $17$ edges. These graphs  were quite different from one another, with the three graphs having only two edges in common, the Wingara and Clare graphs sharing eight edges, and the Willunga graph having three edges in common with the Wingara and Clare graphs. Given the paucity of the data and the complexity of the models, this lack of concordance between the graphs obtained separately for each vineyard is not surprising. 

 In order to make more efficient use of the data,  models  incorporating data from all three vineyards simultaneously were then considered.

The question of how best to include temperature and vineyard effects in the model for gene expression was investigated using linear regression models with forward and backward selection. The largest model fitted for each gene contains separate intercepts for the data from each vineyard, and terms for each of the temperatures recorded $30,90,150,210,270$ and $330$ minutes before the grapes were picked. 
We also considered the model including vineyard and temperature main effects and two-way temperature interactions. For the full model with interactions, it was observed that the adjusted $R^2$ of many of the regressions was above $0.99$, indicating that some over-fitting was  taking place. We  therefore exclude two-way temperature interaction effects in what follows.

Results of the stepAIC function in R, \cite{R}, indicate that there is not a single backwards elimination step that would apply to all genes. That is, each of the vineyard or temperature variables is significant in at least one of the $26$ regression models estimated. In any case, use of separate regression models for each gene is beyond the scope of the present score metrics. As such, the  largest model considered is as follows:
\[
{x}_i \mid {x}_{P_i}, {\gamma}_i, \psi_i, {b}_i \sim N_{50}\left({x}_{P_i} {\gamma}_i + Q{b}_i, \psi_i I_{50}\right), \]
\[
{\gamma}_i \mid \psi_i \sim N_{|P_i|}({0}, \tau^{-1}\psi_i I_{|P_i|}), \]
\[
\psi_i ^{-1}\sim Ga \left( \frac{\delta + | P_i|}{2}, \frac{\tau}{2} \right), \]
where ${b}_i = \left( b_{i1}, \ldots, b_{i9}\right)$ and $b_{ij} \ (j = 1,2, 3)$ is the effect of vineyard $j$ on the expression level of gene $i$, $b_{i4}, and \ldots, b_{i9}$ are the  temperature effects. 

\begin{figure}
\begin{center}
\includegraphics[width=6cm]{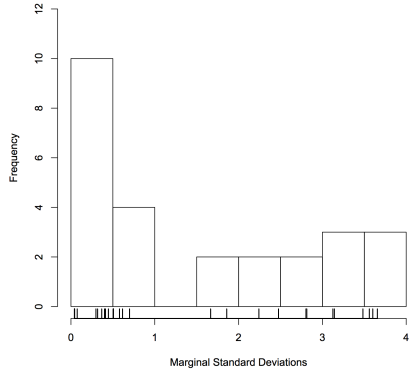}  \\
(a) Standard errors of expression levels\\
\includegraphics[width=6cm]{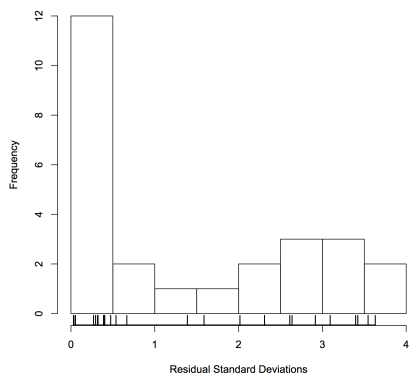} \\
(b) Residual standard errors after regressing expression levels on vineyard\\
\includegraphics[width=6cm]{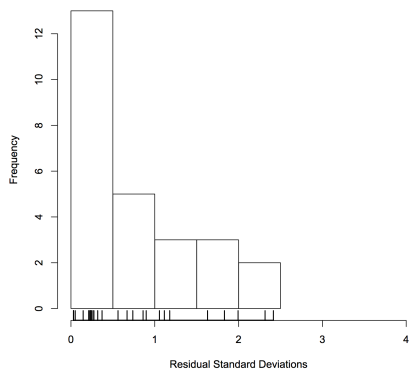} \\ 
	(c)Residual standard errors after regressing expression levels on vineyard and temperature\\
\end{center}
\caption{Histograms of  the marginal standard deviations of the grape gene expression levels and the residual standard errors after regressing the expression levels on vineyard only, then vineyard and temperature.} \label{StDevsFig}

\end{figure}

Histograms of the marginal standard deviations of the expression data for each gene, and the residual standard errors from the regressions containing vineyard, and vineyard and temperature as covariates, are shown in Figure \ref{StDevsFig}. Note that there are three genes with very small standard deviations. These plots show that vineyard variables account for only some of the variation in the gene expression levels. Changes in temperature and vineyard account for much more of the observed variation, but there remains some residual variation to be explained. On the basis of these histograms, we expect that the graph obtained when only vineyard is included as an exogenous variable will be somewhat similar to that obtained when the exogenous variables are ignored, whilst we would expect to see a reasonably different structure when both vineyard and temperature are accounted for.

In accounting for the effects of vineyard and temperature, we find high-scoring Bayesian networks using the BGeCM score metric,  first fitting the model with vineyard effects only, then the model with vineyard and main temperature effects. The highest-scoring Bayesian networks found for $\upsilon = 0.5,1$ and $10$ were recorded and their moral graphs  summarized in Table \ref{GraphTable}. It can be seen that as more covariates are included in the model, more of the variation in the expression levels of the grapes is explained, and the highest-scoring graphs obtained have fewer edges. Edges that are removed as more exogenous variables are included in the model can be interpreted as being explained by common relationships of genes with these additional covariates.

The BGeCM score metric assumes that the effects of the exogenous variables are independent and identically distributed, an assumption that must be questioned. The effects of temperature and vineyard are almost certainly not {\it  iid}. However, there is little information available to provide a useful estimate of the covariance structure of the effects of these exogenous variables. Therefore, the residual approach, which makes no assumptions about the covariance structure of the effects included in the model, is preferred here.

\begin{table}
 \caption{Number of edges in the moral graphs associated with the highest-scoring Bayesian networks for $3$ different sets of included exogenous variables, for the BGeCM score with varying values of $\upsilon$, where $b_i \sim N(0, \upsilon^{-1}\psi_i)$, and for the residual method.}{
\label{GraphTable}

\begin{tabular}{ c  | c  c  c | c  }
&  \multicolumn{3}{c|}{ BGeCM   } &  \\ 
{}& \multicolumn{3}{c|}{${\upsilon}$   } & Residual  \\ 
{Included Covariates} &   $ 0.5 $ & $1 $ & $10 $  &\\  \hline
Vineyard & 66 & 76 & 89 & 68 \\
Vineyard and temperature & 57  & 63 & 68 & 41
 \end{tabular}}
\end{table}

The number of edges in the moralized versions of the highest-scoring networks obtained using the residual method are summarized in Table \ref{GraphTable}.  When only the effects of vineyards are included in the model, the results obtained using the residual method are similar to those obtained when the BGeCM score is used, as expected on the basis of the histograms in Figure \ref{StDevsFig}. When the effects of temperature are included in the model, the residual method produces high-scoring graphs with fewer edges than the BGeCM score metric. This indicates that whilst the BGeCM score metric may account correctly for the covariance structure of the effects of vineyard, the effects of temperature may have a more complicated variance structure, that is not adequately modelled by the {\it  iid} assumption.

The moralized graphs obtained from the residual method are displayed in Figure \ref{GrapeGraphs}.  These graphs, drawn using GraphViz, \cite{graphviz}, show that as more  of the variation in gene expression due to exogenous sources is accounted for in the model, the moral graphs of the highest-scoring networks obtained  have fewer edges. 
The graph obtained by including both temperature and vineyard as exogenous variables, Figure \ref{GrapeGraphs}(c), is preferable to that obtained when only vineyard is included, Figure \ref{GrapeGraphs}(b). For most genes, very little variation in the gene expression values is accounted for by the relationship with vineyard alone. 

There are a number of interesting features to be observed in  graph  Figure \ref{GrapeGraphs}(c), which is the graph obtained when both vineyard and temperature effects are accounted for in the model.  We observe that seven nodes in this graph are completely disconnected from all other nodes, which implies that once relationships with temperature and vineyard have been accounted for, the expression levels of each of these genes are independent of the expression levels of all other genes.  (Recall that absence of an edge between two genes in Figure \ref{GrapeGraphs}(c) indicates that the expression levels of these nodes are independent, a relationship which can be refined through application of Markov properties.)

It is apparent that three of these seven disconnected nodes, corresponding to genes $14$, $18$ and $23$, are already disconnected from the rest of the graph when only vineyard  is included in the model; see Figure \ref{GrapeGraphs}(b), where it is observed that these are the only three unconnected nodes.  The expression levels of these three genes are observed to have the lowest standard deviations of all the genes, at  $0.037, 0.034$ and $0.068$ respectively, and are in fact the three genes at the left-most end of the rug in Figure \ref{StDevsFig}(a). When these three genes are regressed on vineyard, the residual standard deviations are even smaller ($0.029, 0.023$ and $0.047$).  In other words, there is no variation in the expression levels of these genes to be explained by relationships with other genes, so it is not surprising that they are unconnected in the graph.  Very small gene standard deviations can be problematic in microarray data analysis, and methods have been proposed for adjusting the standard deviation estimates upwards by adding a constant term or by application of empirical Bayes methods when constructing $t$-tests, for example, \cite{Smyth:2004aa}.  Such adjustment  is beyond the scope of our present analysis however. Note that the fact the three genes are connected in Figure \ref{GrapeGraphs}(a) is suggestive of overfitting when the BGe metric is used.

Genes  $9$, $10$ and $11$ are also disconnected in the final graph,  Figure \ref{GrapeGraphs}(c), and correspond to Hsp81, which  is an early response to dehydration. According  to the KEGG data base, \cite{Kanehisa:2010aa}, they are predicted to be similar to Hsp90.   The role of these sets of  genes in the heat shock network of grapes is not entirely understood. The role of Hsp81 proteins in \emph{Arabidopsis thaliana}, more commonly known as thale cress, has been discussed in \cite{Yabe}, who note that an increase in the expression level of Hsp81-1 is possibly caused by a regulatory pathway other than the heat shock pathway.  Our analysis supports this finding for \emph{Vitis Vinifera}, indicating that Hsp81 may not be implicated in the heat shock protein network of grapes, at least over the four-week time period studied.  We have established that variation in the Hsp81 genes is accounted for directly by the effects of the exogenous variables, and that they are uncorrelated with the other HSPs in the final graph.

The seventh gene, number $26$, which is a mitochondrial small Hsp, is not implicated in the final network either.  This gene is the only mitochondrial gene considered. That it is unconnected from the rest of the network indicates that variation in this gene is explained purely by exogenous temperature and vineyard effects, and is not dependent upon any of the other genes in the dataset.  This suggests that the mitochondrial HSP are not regulated in the same way as other cellular HSPs.

On the whole, relatively little is known about the heat shock regulatory network for grapes. Typically in the representation of the heat response network for plants, relationships between classes of genes, such as small Hsps or Hsp70s are discussed, \cite{HSPs}. The graph obtained here  provides a good starting point for the development of a finer structure, which can then  be further developed. The edges between the genes in Figure \ref{GrapeGraphs}(c) can be interpreted as encoding conditional dependence relationships. This graph is the moralized version of the directed acyclic graph found, and more detail is available through consideration of the class partially directed acyclic graph, or class PDAG, of the underlying directed acyclic graph.   However, since we are analysing observational data, we will consider  here edges in the moralized version of the graph only.  We observe that there are two pairs of genes connected by a single edge. The first  pair of nodes is $(12,19)$, representing   a chaperone gene (gene $12$) which promotes the folding and unfolding of proteins and a class I small HSP (gene $19$).  This is an undirected edge, but the two genes are correlated after adjusting for the exogenous variables, and the chaperone gene $12$ has no other connecting edges.   The second pair of connected nodes is $(2,8)$, representing a glucose regulated HSP (gene $2$) and the ubiquitin conjugating enzyme 4e (gene $8$); again these two genes are correlated after adjusting for the exogenous variables, and the enzyme gene $8$ has no other connecting edges.

Further investigation of the connected nodes and possible regulatory heat shock mechanisms is beyond the scope of the present paper, and would require further biological evidence and possible investigation.  It is clear from the detection of the unconnected nodes together with the plausible relationships between nodes connected by a single edge, that  analysis of the data using the new score metrics  has demonstrable utility and has detected real biological effects.

\section{Discussion}\label{ConclusionSection}

The BGeCM score metric and the residual approach presented in this paper enable Bayesian network structures to be learnt given  datasets that do not consist of independent and identically distributed samples, and may be used in conjunction with any score-based method for the estimation of a Bayesian network. 
Furthermore, the residual approach allows the estimation of a Bayesian network for datasets with a complex mean structure without the need to specify the variance structure of the mean effects. This approach proved  useful for the analysis of the grape-berry gene data, where it could not reasonably be supposed that the effects of the exogenous variables were independent and identically distributed.
Our analysis of the grape-berry gene microarray data has resulted in biologically plausible conclusions on the heat shock regulatory network of grape genes. These inferences could not have been drawn without the availability of suitable score metrics  to account for the effects of exogenous variables.

\bibliography{PhDReferences}

\begin{thebibliography}{10}

\bibitem{CoombeGrape}
B.~G. Coombe.
\newblock The regulation of set and development of the grape berry.
\newblock {\em Acta Horticulturae}, 34:261--271, 1973.

\bibitem{Davison}
A.~C. Davison.
\newblock {\em Statistical Models}.
\newblock Cambridge Series in Statistical and Probabilistic Mathematics.
  Cambridge University Press, 2003.

\bibitem{HiddeJong}
H.~de~Jong.
\newblock Modeling and simulation of genetic regulatory systems: A literature
  review.
\newblock {\em Journal of Computational Biology}, 9(1):67--103, 2002.

\bibitem{SAMSI}
A.~Dobra, C.~Hans, B.~Jones, J.R. Nevins, and M.~West.
\newblock Sparse graphical models for exploring gene expression data.
\newblock {\em Journal of Multivariate Analysis}, 90(1):196--212, 2004.

\bibitem{Dykstra}
R.~L. Dykstra.
\newblock Establishing the positive definiteness of the sample covariance
  matrix.
\newblock {\em Annals of Mathematical Statistics}, 41(6):2153--2154, 1970.

\bibitem{graphviz}
J.~Ellson, E.~R. Gansner, E.~Koutsofios, S.~C. North, and G.~Woodhull.
\newblock Graph drawing software.
\newblock chapter Graphviz and dynagraph - static and dynamic graph drawing
  tools, pages 127--148. Springer-Verlag, 2004.

\bibitem{FriedmanScience}
N.~Friedman.
\newblock Inferring cellular networks using probabilistic graphical models.
\newblock {\em Science}, 303:799--805, February 2004.

\bibitem{Friedman}
N.~Friedman, M.~Linial, I.~Nachman, and D.~Pe'er.
\newblock Using {B}ayesian networks to analyze expression data.
\newblock {\em Journal of Computational Biology}, 7:601--620, 2000.

\bibitem{LearningGaussianNetworks}
D.~Geiger and D.~Heckerman.
\newblock Learning {G}aussian networks.
\newblock In {\em Proceedings of the Tenth Conference on Uncertainty in
  Artificial Intelligence}, 1994.

\bibitem{GeigerHeckerman}
D.~Geiger and D.~Heckerman.
\newblock Parameter priors for directed acyclic graphical models and the
  characterization of several probability distributions.
\newblock {\em The Annals of Statistics}, 30(5):1412--1440, October 2002.

\bibitem{Irizarry}
R.~A. Irizarry, B.~Hobbs, F.~Collin, Y.~D. Beazer-Barclay, K.~J. Antonellis,
  U.~Scherf, and T.~P. Speed.
\newblock Exploration, normalization, and summaries of high density
  oligonucleotide array probe level data.
\newblock {\em Biostatistics}, 4(2):249--264, 2003.

\bibitem{Kanehisa:2010aa}
M.~Kanehisa, S.~Goto, M.~Furumichi, M.~Tanabe, and M.~Hirakawa.
\newblock {KEGG} for representation and analysis of molecular networks
  involving diseases and drugs.
\newblock {\em Nucleic acids research}, 38:D355--D360, 2010.

\bibitem{KaszaSolomon}
J.~Kasza and P.~Solomon.
\newblock Kullback {L}eibler divergence for {B}ayesian networks with complex
  mean structure.
\newblock arXiv:1009.1463.

\bibitem{Kasza}
J.~E. Kasza.
\newblock {\em {B}ayesian networks for high-dimensional data with complex mean
  structure}.
\newblock PhD thesis, The University of Adelaide, 2009.

\bibitem{KollerFriedman}
D.~Koller and N.~Friedman.
\newblock {\em Probabilistic graphical models}.
\newblock MIT Press, 2009.

\bibitem{KotakHSP}
S.~Kotak, J.~Larkindale, U.~Lee, P.~{von} Koskull-D\"{o}ring, E.~Vierling, and
  K.-D. Scharf.
\newblock Complexity of the heat stress response in plants.
\newblock {\em Current opinion in plant biology}, 10:310--316, 2007.

\bibitem{Lauritzen}
S.~L. Lauritzen.
\newblock {\em Graphical Models}.
\newblock Clarendon Press, Oxford, 2004.

\bibitem{Markowetz}
F.~Markowetz and R.~Spang.
\newblock Inferring cellular networks - a review.
\newblock {\em BMC Bioinformatics}, 8(Suppl 6):S5, September 2007.

\bibitem{DaviesGrape}
S.~P. Robinson and C.~Davies.
\newblock Molecular biology of grape berry ripening.
\newblock {\em Australian Journal of Grape and Wine Research}, 6:175--188,
  2000.

\bibitem{Smyth:2004aa}
G.~K. Smyth.
\newblock Linear models and empirical {B}ayes methods for assessing
  differential expression in microarray experiments.
\newblock {\em Statistical Applications in Genetics and Molecular Biology},
  3(1), 2004.

\bibitem{Speed:1997aa}
T.~P. Speed.
\newblock {\em Encyclopedia of Statistical Sciences Update Volume 1}, chapter
  Restricted maximum likelihood ({REML}).
\newblock Wiley, 1997.

\bibitem{SGS}
P.~Spirtes, C.~Glymour, and R.~Scheines.
\newblock {\em Causation, Prediction, and Search}.
\newblock Springer-Verlag, 1993.

\bibitem{R}
R~Development~Core Team.
\newblock {\em {R}: A Language and Environment for Statistical Computing}.
\newblock R Foundation for Statistical Computing, Vienna, Austria, 2007.

\bibitem{HSPs}
W.~Wang, B.~Vinocur, O.~Shoseyov, and A.~Altman.
\newblock Role of plant heat-shock proteins and molecular chaperones in the
  abiotic stress response.
\newblock {\em Trends in Plant Science}, 9(5):244--252, May 2004.

\bibitem{Yabe}
N.~Yabe, T.~Takahashi, and Y.~Komeda.
\newblock Analysis of tissue-specific expression of \emph{Arabidopsis thaliana}
  {HSP90}-family gene {HSP81}.
\newblock {\em Plant cell physiology}, 35(8):1207--1219, 1994.

\end{thebibliography}
\bibliographystyle{plain}

\end{document}